# Illustrating Special-Relativity Phenomena via Gaussian Ray Optics

M. A. Bouchene

*Laboratoire Collisions Agrégats Réactivité, FeRMI, Université de Toulouse and CNRS UMR 5589, Toulouse, France*

**Abstract**

We highlight the correspondence between one-dimensional Lorentz transformations, which relate events observed from two distinct inertial reference frames, and ray transfer transformations in Gaussian optics. Specifically, we identify optical systems whose transfer matrices reproduce the mathematical structure of the Lorentz transformation. Within this framework, we show that fundamental effects of special relativity as time dilatation, length contraction and loss of synchronicity find a direct counterpart in the behavior of light rays deviated by a diverging lens, thereby providing a novel and illustrative optical equivalence of relativistic phenomena. This article is intended for a broad audience of physicists. The correspondence described is general and may interest students with a moderate background in special relativity and optics, as well as teachers in the corresponding fields.

# I- Introduction

The well-known adage *"money has no smell"* could just as well be applied to mathematical theories. Indeed, it is often the case that the same mathematical formalisms and tools can describe entirely different physical situations. For instance, tensor calculus appears both in general relativity and in fluid mechanics, while probability theory is equally at home in statistical physics and quantum mechanics.

Special Relativity (SR) and Gaussian Optics (GO) share a common feature, linearity, meaning a linear relationship between the coordinates of events (in SR) or ray coordinates (in GO) at the input and output. This linearity can be cast in a matrix relationship between these quantities. In a one-dimensional spatial context, both SR and paraxial geometric optics use 2×2 matrices: Lorentz transformation matrices in SR and ray transfer matrices in GO for centered optical systems.

Although this connection between two seemingly unrelated fields might appear superficial, it proves to be fruitful when the correspondence is pursued further. In this article, we develop this approach by explicitly determining the optical system that mimics a given Lorentz transformation. We then explore the equivalence between some of the most well-known SR phenomena—such as time dilation, length contraction, and relativity of simultaneity—and the corresponding behavior of light rays propagating through such an optical system.

This example provides both a compelling illustration of SR and a demonstration of how an abstract concept—linearity—can bridge different fields that rely on the same mathematical tool ; the matrix formalism. The resulting correspondence not only can be useful for better illustration of SR but may also inspire further theoretical developments.

# I- One-dimensional special relativity :

The Lorentz transformations (LT) are fundamental equations in SR that describe how space and time coordinates change between two inertial reference frames (R) and (R') (Fig. 1-a). The x-axis of both frames are oriented in the direction of the relative velocity of (R') with respect to (R). These transformations can be derived from the invariance of the speed of light c (Einstein's historical approach) [5], the invariance of the space-time interval $s^2 = c^2t^2 - x^2$

[6] or from more general hypotheses such as the homogeneity of space-time combined with the principle of causality [7]. The latter approach is particularly insightful, as it justifies the required linearity of the Lorentz transformations.

Let $(x, ct)$ and $(x', ct')$ denote the space-time coordinates of an event in frames R and R' respectively and $(\Delta x, c\Delta t), (\Delta x', c\Delta t')$ the corresponding space-time intervals between two events. The Lorentz transformation then takes the form $\begin{pmatrix} x \\ ct \end{pmatrix} = M_{SR} \begin{pmatrix} x' \\ ct' \end{pmatrix}$ and $\begin{pmatrix} \Delta x \\ c\Delta t \end{pmatrix} = M_{SR} \begin{pmatrix} \Delta x' \\ c\Delta t' \end{pmatrix}$ respectively with :

$$M_{SR} = \begin{pmatrix} \gamma & \gamma\beta \\ \gamma\beta & \gamma \end{pmatrix} \qquad (1)$$

The Lorentz transformation matrix (LTM) in SR , $\beta = u/c$ and $\gamma = \dfrac{1}{\sqrt{1-\beta^2}}$ the Lorentz factor. $u$ is the relative velocity of (R') with respect to (R) and our choice of the orientation of x-axis implies that $\beta > 0$.

A key property of the Lorentz matrix—stemming from the invariance of the space-time interval $s^2 = c^2 t^2 - x^2$ is that its determinant remains constant regardless of the relative velocity, specifically $\begin{vmatrix} \gamma & \gamma\beta \\ \gamma\beta & \gamma \end{vmatrix} = 1$

The properties of this simple matrix, especially its symmetry, account for all kinematic effects in one-dimensional SR, including those that appear paradoxical.

## II- <u>Gaussian optics:</u>

Gaussian optics is a fundamental approximation in both geometrical and wave optics for centred optical systems. It relies on the assumption that light rays are paraxial**,** meaning they stay close to the optical axis and form small angles with it [3, 4, 8].

One of the most powerful mathematical tools in paraxial optics is the use of transfer matrices commonly referred as « ABCD » matrices. These matrices provide a compact and efficient way to describe how an optical system transforms an incoming light beam. The « ABCD » matrix formalism is especially useful for analyzing the propagation of a Gaussian beam and assessing the stability of optical resonators. Each optical elements -such as lenses, mirrors, and free-space propagation- is represented by a 2×2 matrix. The overall system is described by the product of these matrices.

Consider a centred optical system characterized by a transfer matrix $\begin{pmatrix} A & B \\ C & D \end{pmatrix}$. A light ray enters the system through an input plane (and exits at an output plane), located at distances $r$, $r'$ respectively from the optical axis. The ray propagates at small angles $\alpha$ and $\alpha'$ with respect to the axis. The relationship between the input and output parameters is given by:

$$\begin{pmatrix} r \\ n\alpha \end{pmatrix} = \begin{pmatrix} A & B \\ C & D \end{pmatrix} \begin{pmatrix} r' \\ n'\alpha' \end{pmatrix} \tag{2}$$

For simplicity, we consider that the propagation in both the input and output media takes place in vacuum (i.e. refraction indexes $n = n' = 1$). An important result is that the determinant of the transfer matrix is equal to 1, i.e. $AB - CD = 1$. This is the consequence of the conservation of phase-space area (or etendue) in paraxial optics [8].

In eqt. 2, $A$ and $D$ are dimensionless while $B$ has the dimension of a length and $C$ the dimension of the inverse of a length. By introducing an arbitrary but fixed unit of length $r_0$, we can rewrite this relationship in the form:

$$\begin{pmatrix} r/r_0 \\ n\alpha \end{pmatrix} = \begin{pmatrix} A & B/r_0 \\ C r_0 & D \end{pmatrix} \begin{pmatrix} r'/r_0 \\ n'\alpha' \end{pmatrix} \tag{3}$$

In this relation, the matrix elements are all dimensionless and correspondence with the Lorentz transformation can be done properly.

### III- Correspondence

The correspondence between the two domains can be emphasized by equating the two matrices LTM (see eqt. 1) and GM (see eqt. 3). The LTM depends solely on a single parameter, $\beta$, but its symmetric structure imposes a strong constraint on its form. Whatever is the LTM, we always can find an optical system that exhibits an equivalent transfer matrix. Indeed, equating the two matrices leads to the corresponding optical coefficients :

$$A = D = \gamma \qquad (4)$$
$$B = C r_0^2 = \gamma \beta r_0$$

Conversely, if we consider an optical system with a given transfer matrix, the equivalent LTM can not always exists since in addition to the relations (4), the following constraints have to be fullfilled : $\gamma \geq 1, \beta \geq 0$, leading to $A = D \geq 1$ and $B = C r_0^2 > 0$. When done, the relation $\gamma = 1/\sqrt{1-\beta^2}$ is automatically fullfilled from the equality $AB - CD = 1$. Note that the unit of length is arbitrary but fixed, and all associated distances must use the same unit, allowing for an infinite number of equivalent optical systems that differ only in scale.

We consider next some concrete cases where correpondence can be established and show how SR phenomena can be visualized through simple optical effects. For simplicity, we set throughout the rest of the paper $r_0 = 1$.

### III-1 Case of an optical cavity

The constraint $AB - CD = 1$ implies that the transfer matrix depends on only three independent parameters. Furthermore, any such matrix is equivalent to the transfer matrix of an optical cavity composed of two mirrors, $M_1$ and $M_2$ with radii of curvature $R_1$ and $R_2$, separated by a distance $d$ (i.e three parameters also). If we define the input and output planes as lying just before mirror $M_1$, and consider light propagating from $M_1$ to $M_2$, then the transfer matrix $M_T$ of the cavity is given by the product of individual matrices:
$\begin{pmatrix} 1 & d \\ 0 & 1 \end{pmatrix} \begin{pmatrix} 1 & 0 \\ -2/R_2 & 1 \end{pmatrix} \begin{pmatrix} 1 & d \\ 0 & 1 \end{pmatrix} \begin{pmatrix} 1 & 0 \\ -2/R_1 & 1 \end{pmatrix}$. This leads to the final result for the Gaussian matrix (GM):

$$M_T = \begin{pmatrix} (1-2d/R_2)(1-2d/R_1) - 2d/R_1 & 2d(1-d/R_2) \\ (-2/R_2)(1-2d/R_1) - 2/R_1 & 1-2d/R_2 \end{pmatrix} \qquad (5)$$

The correspondence between the two domains can be emphasized by equating the two matrices LTM (see eqt. 1) and GM (see eqt. 5) or equivalently by using relations (4) with $r_0 = 1$. Performing a straightforward algebraic calculation yields the following expressions for the parameters $d$, $R_1$ and $R_2$:

$$R_1 = \infty \qquad (6\text{-a})$$

$$R_2 = -2/\gamma\beta \qquad (6\text{-b})$$

$$d = (\gamma - 1)/\gamma\beta \qquad (6\text{-c})$$

The corresponding optical resonator consists of a plane mirror $M_1$ and a convex mirror $M_2$ with $R_2 < 0$, as shown in Fig. 1-b.

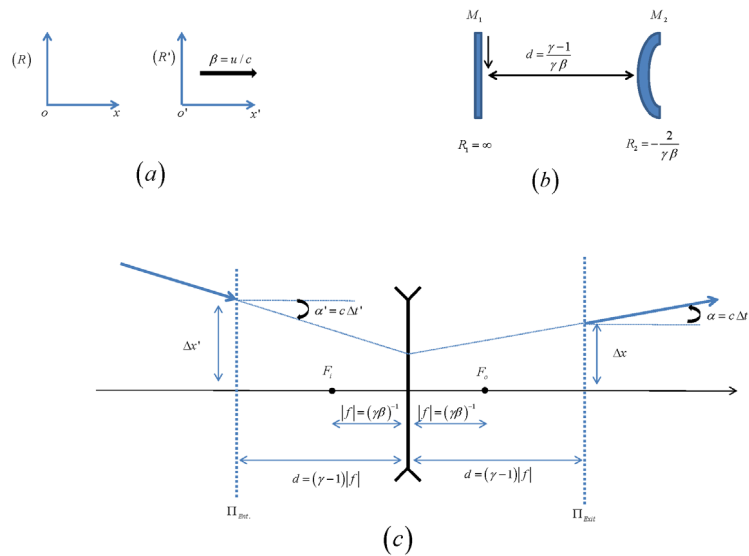

### III-2 Case of a diverging lens

Another simple optical system can be found by "unfolding" the cavity into a linear optical system (Fig. 1-c), obtaining an equivalent setup: a homogeneous, empty medium (vacuum) of length d, followed by a diverging lens with focal length $f = -(\gamma\beta)^{-1}$, and then another section of vacuum of length $d = (\gamma - 1)/\gamma\beta$.

**Figure 1** : Optical equivalence

(a) *The two inertial referentials.* The x-axis of frames (R) and (R') are assumed to be oriented in the same direction as the relative velocity of (R') with respect to (R) (i.e. $\beta > 0$)

(b) *The optical resonator equivalent to the Lorentz transformation.* The diagram shows an optical resonator whose transfer matrix is formally equivalent to the Lorentz transformation between the two reference frames. The vertical arrow indicates the input/output planes of the optical system. Mirror $M_1$ is flat, while mirror $M_2$ is convex.

(c) *Expanded representation of the optical cavity.* This extended diagram illustrates the ray propagation within the equivalent optical system. A light ray enters the system on the input plane ($\Pi_{Ent.}$), at a point located at a distance $\Delta x'$ from the optical axis and inclined at an angle $\alpha' = c\Delta t'$. After passing through the divergent lens, the ray exits on the output plane ($\Pi_{Exit}$), at a point located at a distance $\Delta x$ from the axis and with an inclination $\alpha = c\Delta t$. The focal length of the lens and the relevant geometrical distances are also shown and expressed as functions of the relativistic parameters.

This optical equivalence allows for a reinterpretation of the LT: a light ray, with coordinates $(x, ct)$ at the input plane, propagates through the equivalent optical system and exits with coordinates $(x', ct')$. Considering intervals of time and space, we denote $(\Delta x, c\Delta t)$ and $(\Delta x', c\Delta t')$ accordingly.

An important consequence of this correpondence is the geometric interpretation of simultaneity and co-location:

- Two events that are simultaneous in frame (R′) $(\Delta x', c\Delta t' = 0)$ or in frame (R) $(\Delta x, c\Delta t = 0)$ are represented by a light ray that is parallel to the optical axis at the input or output planes, respectively.
- Two events that occur at the same spatial location in frame (R′) $(\Delta x' = 0, c\Delta t')$ or in frame (R) $(\Delta x = 0, c\Delta t)'$ are represented by a light ray intersecting the optical axis at the input or output plane, respectively.

The simplicity of the equivalent optical system makes it possible to grasp fundamental concepts of SR intuitively, particularly through the analysis of three special cases. Figures 2 to 5 illustrate the case where $d > f$, i.e., $\gamma > 2$. The geometric constructions for the case $d \leq f$ (i.e. $\gamma \leq 2$) lead to the same conclusions. We leave it to the reader to verify this configuration.

*III-2-a- Length contraction*:

Starting from the LT, expressed via the transformation matrix (see Eq. 1), it can be readily shown that in SR, two events that occur simultaneously in frame (R) (i.e. $\Delta t = 0$), but are spatially separated in frame (R′) by a distance $\Delta x' \neq 0$, are perceived by an observer in frame (R) as being separated by a shorter distance $\Delta x = \Delta x'/\gamma$.

This corresponds to the scenario where one measures, in frame (R), the length $L$ of an object whose proper length is $L_0$ in its rest frame (R′). It is important to note that such a measurement requires a simultaneous determination—in frame (R)—of the positions of the object's moving endpoints (i.e., both measured at the same time $t$ in (R)). As a result, the relation $L = L_0/\gamma$ holds, demonstrating the length contraction effect in SR.

In the optical equivalence (see Figure 2), this situation is modeled by a light ray striking the input plane ($\Pi_{ent}$) at a distance $\Delta x' = L_0$ from the optical axis. The condition of simultaneous measurement in

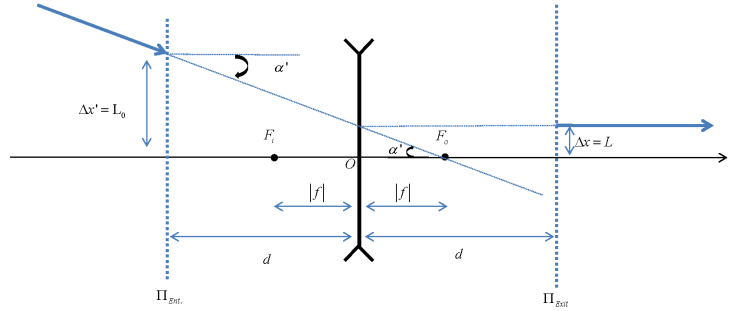

Figure 2 : Length contraction

Measurement of the length of an object with proper length $L_0$ (in R'). Synchronous measurement of the positions of its endpoints in (R) forces the optical ray at the output plane ($\Pi_{Exit}$) to be parralel to the optical axis. Elementary geometrical analysis then allows us to determine the correponding contracted length $\Delta x = L_0 / \gamma$ in (R) (see main text for derivation).

frame (R) corresponds to the transmitted ray being parallel to the optical axis. According to the laws of geometrical optics, this condition is met only if the incident ray makes an angle $\alpha'$ such that its extension passes through the focal point $F_0$ of the diverging lens.

Considering the diverging nature of the lens, the resulting ray geometry clearly illustrates that the image height is reduced compared to the object height, i.e. $\Delta x < \Delta x'$. Moreover, a straightforward geometric analysis allows us to derive the exact law of length contraction. Applying Thales's theorem yields the relation: $L / L_0 = |f| / (d + |f|)$. Combining this with Eq. (6-c), and using the expression for the focal length $f = -1 / \gamma\beta$ of the equivalent optical system, we recover the standard Lorentz contraction formula: $L / L_0 = 1 / \gamma$.

*III-2-b- Time dilatation*:

Starting from the LT, expressed via the transformation matrix (see Eq. 1), it can be readily shown that in SR, two events that occur at the *same spatial location* in frame (R') (i.e., $\Delta x' = 0$) and are separated by a time interval $\Delta t'$, are observed from the perspective of an observer in frame (R) as being separated by a *longer* time interval $\Delta t$, such that: $\Delta t = \gamma \, \Delta t'$. This illustrates the time dilation effect, where moving clocks appear to run slower when observed from a different inertial frame.

To illustrate this using the optical equivalence (see Figure 3), consider a light ray that enters the system through the input plane exactly along the optical axis at point (A), but with an inclination angle $\alpha' = c\Delta t'$. The ray propagates without deviation through the homogeneous medium until it reaches the diverging lens, where it is refracted. Due to the divergent nature of the lens, the output angle $\alpha = c\Delta t$ is necessarily larger than the input angle $\alpha' = c\Delta t'$,

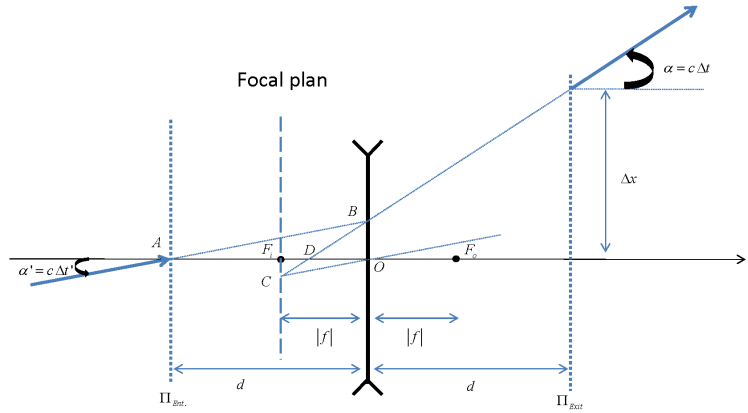

Figure 3 : Time dilatation

Two events separated by time interval $\Delta t'$ in (R') and occuring at the same spatial position (i.e. $\Delta x' = 0$) are represented by a lightl ray that strikes the entrance plane ($\Pi_{Ent.}$) at point A located on the optical axis. The corresponding time interval $\Delta t$ in frame (R) is related to the inclination $\alpha = c\Delta t$ of the emerging ray. Geometrical analysis allows to derive the relationship $\Delta t = \gamma \Delta t'$ (see main text for details).

directly reflecting the time dilation phenomenon: the "temporal separation" represented by the output angle is larger.

A straightforward geometric analysis yields the desired quantitative relation. From Figure 3, the following relations hold: $OB = OD\alpha = d\alpha'$, $F_iC = |f|\alpha'$. Applying Thales's theorem gives: $F_iC/OB = (|f| - OD)/OD$. Combining these expressions leads to: $\alpha/\alpha' = (d + |f|)/|f|$. Then, using Eq. (6-c) and the known expression for the focal length $f = 1/\gamma\beta$ of the system, we recover the time dilation formula: $\alpha/\alpha' = \Delta t/\Delta t' = \gamma$.

*III-2-c- Loss of synchronicity:*

Another important consequence of SR is the loss of simultaneity between events when changing reference frames. From the LT, it follows that if two events are simultaneous in a given frame (R') (i.e., $\Delta t' = 0$) but occur at different spatial locations ($\Delta x' \neq 0$), they will be observed in frame (R) as being separated by a nonzero time interval ($\Delta t \neq 0$); in other words, they are no longer simultaneous. Moreover, elementary algebraic calculations gives the relation $\Delta t = \gamma \beta \Delta x'/c$.

This is perhaps one of the most counterintuitive consequences of SR, as it implies that the concepts of past, present, and future are relative—they depend on the observer's state of motion.

This effect is illustrated in Figure (4). A light ray enters the optical system at the input plane ($\Pi_{ent}$) at a $\Delta x'$ from the optical axis and travels parallel to the axis, representing two simultaneous events in frame (R′) (here, $\Delta x' > 0$ for simplicity). Upon passing through

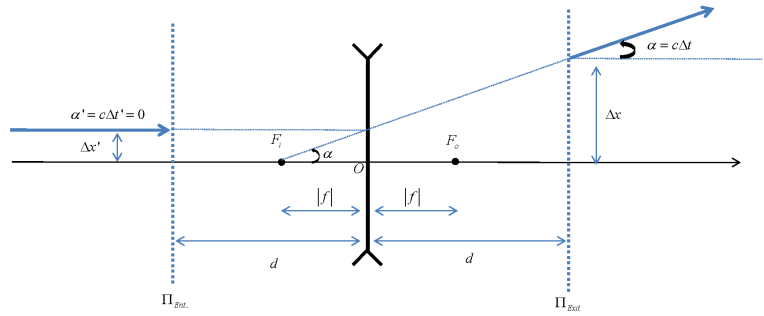

**Figure 4** : Loss of synchronicity

Two synchronous events in (R') and occuring at different spatial positions are represented by an light ray entering the system parrallel to the optical axis and striking the entrance plane ($\Pi_{Ent.}$) at a distance $\Delta x'$ from te axis. The ray emerges from the output plane ($\Pi_{Exit}$) with a non zero inclination angle $\alpha = c\Delta t$ provided that $\Delta x' \neq 0$. The two events although simultaneous in (R') are no longer synchronous in (R).

the diverging lens, the ray exits the system at plane ($\Pi_{exit}$) with an inclined trajectory, since it must pass through the focal image point $F_i$. This inclination reflects the fact that the two events are no longer simultaneous in frame (R), except in the special case where the input ray lies exactly along the optical axis ($\Delta x' = 0$).

The corresponding time interval $\Delta t$ between the events in frame (R) can be deduced geometrically from the figure. In particular: $\alpha = \Delta x'/|f|$ and thus $\Delta t = \gamma \beta \Delta x'/c$.

Furthermore, the geometric construction reveals that, unlike the case of length contraction, the two events are perceived in frame (R) as having a spatial separation $\Delta x > \Delta x'$. This is a useful example to help students distinguish between the spatial separation of events and the measurement of the length of an object, which requires a simultaneous measurement of the object's endpoints in the same reference frame—as previously discussed in Section III-2-a.

*III-2-d- Velocity additivity*

Let us consider an object moving relative to two inertial reference frames, (R) and (R′), with velocities $V$ and $V'$, respectively. In frame (R′), the object travels a distance $\Delta x' = V'\Delta t'$ during a time interval $\Delta t'$. In frame (R), an observer measures a corresponding distance $\Delta x = V \Delta t$ and time interval $\Delta t$. Using the LT, one can derive -through a straightforward algebraic calculation- the relationship between the two velocities $V$ and $V'$: $V = \dfrac{V' + \beta c}{1 + V'\beta/c}$. This is the well-known relativistic velocity addition formula.

In the optical equivalence (Figure 5), we consider a light ray entering the system at point (A) on the input plane $(\Pi_{ent})$, and exiting at point G on the output plane $(\Pi_{exit})$. The ray simulates the motion of the object from one point to another in space-time. The velocity $V$, defined as $V = c\Delta x/\alpha$ can be determined geometrically.

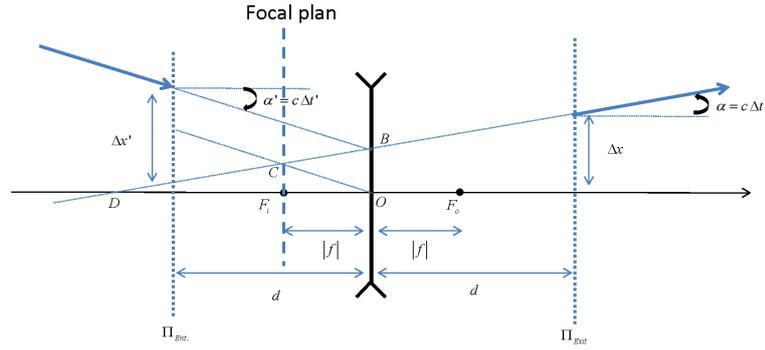

**Figure 5** : Velocity additivity

Two events in (R') correponding to a moving object covering a distance $\Delta x'$ during a time interval $\Delta t'$ are represented by a light ray (solid arrow) that strikes the system at the entrance plane ($\Pi_{Ent.}$). The ray exits the system at the plane ($\Pi_{Exit}$) at a distance $\Delta x$ from the axis and with an inclination angle $\alpha = c\Delta t$. The corresponding velocities in (R') and (R) are $V' = \Delta x'/\Delta t'$ and $V = \Delta x/\Delta t$ respectively. Geometric considerations lead to the relation $V = \dfrac{V' + \beta c}{1 + V'\beta/c}$ (see main text for derivation)

Applying Thales's theorem in the geometric configuration yields: $F_i C/OB = (OD - |f|)/OD$

In this setup, we have $OB = \Delta x' + \alpha' d$ ($\alpha' < 0$ in our scheme) and $F_i C = -\alpha'|f|$. This gives:

$OD = (\Delta x' + \alpha' d)|f|/(\Delta x' + \alpha'(d + |f|))$. We also have the relation: $\Delta x = (d + OD)\alpha$.

Combining these two last equations, we get $V = c\Delta x/\alpha = \left[V' + \left((d^2 + 2d|f|)/(|f| + d)\right)\right]/\left[1 + V'/(|f| + d)\right]$. By substituting the relativistic expressions for $|f|$ and $d$ into the equation, we recover the relativistic velocity transformation formula: $V = \dfrac{V' + \beta c}{1 + V'\beta/c}$.

**IV-    Conclusion** :

We have shown in this paper that the kinematic transformations of SR can be simply illustrated through correspondence with GO. In particular, the propagation of a light ray through a simple optical system—comprising free space and a divergent lens—can effectively illustrate all the fundamental phenomena of SR. This correspondence, grounded in elementary geometric principles, allows relativistic phenomena to be illustrated and interpreted through simple concepts and experimentally accessible and familiar optical setups.  Furthermore, this input–output formalism, based on linear transformations, extends beyond optics to a variety of physical systems—for example, electrical circuits, where ABCD parameters (also known as transmission or chain parameters) are used to model two-port networks [9]—thereby underscoring its broader educational value.